# Optimization of Processing Allocation in Vehicular Edge Cloud based Architecture

*Amal A. Alahmadi, T. E. H. El-Gorashi, and Jaafar M. H. Elmirghani*

*Abstract*— Vehicular edge computing is a new distributed processing architecture that exploits the revolution in the processing capabilities of vehicles to provide energy efficient services and low delay for Internet of Things (IoT)-based systems. Edge computing relies on a set of distributed processing nodes (i.e. vehicles) that are located close to the end user. In this paper, we consider a vehicular edge cloud (VEC) consisting of a set of vehicle clusters that form a temporal vehicular cloud by combining their computational resources in the cluster. We tackle the problem of processing allocation in the proposed vehicular edge architecture by developing a Mixed Integer Linear Programming (MILP) optimization model that jointly minimizes power consumption, propagation delay, and queuing delay. The results show that the closer the processing node (PN) is to the access point (AP), the lower the power consumption and delay, as the distance and number of hops affect the propagation delay and queuing delay. However, the queuing delay at the AP becomes a limiting factor when it operates at a low service rate compared to the traffic arrival rate. Thus, processing tasks at the vehicular nodes (VN) was avoided whenever the objective function included queueing delay and the AP operated at a low service rate. Increase in the AP service rate results in a lower queuing delay and better VN utilization.

## I. INTRODUCTION

The rapid expansion of Internet of Thing (IoT) technologies, driven by smart cities for example, has led to exponential growth in the number of connected devices and to an associated increase in the traffic load [1]. This calls for a new paradigm where the IoT-based requests are processed at the edge of the network to mitigate the pressure of the traffic load on the network infrastructure and, to reduce the delay of the services provided. One of the advantages of edge computing is its ability to reduce both the power consumption and the delay experienced by traffic and processing. This can enhance the energy efficiency of the network and can also improve the quality of the delivered services. With the growth in the number of connected devices and the exponential rise in real-time based traffic, the focus needs to be put on optimizing processing and routing in edge-based architectures to minimize both delay and power consumption. On the other hand, the recent growth in processing capabilities of vehicles, offers a revolutionary vision of vehicle-based processing units as part of the edge computing framework. Vehicular Edge Clouds (VEC) can help realize this vision where a group of vehicles in a car park, at a charging station or at a road traffic intersection, cluster and form a temporary vehicular cloud by combining their computational resources in the cluster.

Several studies have been conducted to investigate energy efficiency in networks and cloud data centers [2]–[5]. A number of studies proposed and optimized data centers physical designs and topologies to minimize the consumed power [6]–[10], and other new decentralized designs such as distributed data centers [11]–[13], and fog/edge data centers [14]. Different techniques were studied in the optimization models, including virtualization [15], [16], network coding [17], [18], content distribution [19], [20], big data processing [21], [22], and the efficient use of renewable energy in these network and cloud systems [23]. Part of our previous efforts were directed to the optimization and modeling of vehicular clouds to achieve energy efficient processing. We proposed our first VEC based architecture in [24] where we compared the power consumed in processing data in central data centers to the power consumed in the vehicular cloud. In [25]–[27], we included the end-to-end architecture in the power minimization problem considering the potential of splitting tasks between the available processors. The software matching problem was investigated as well in [28], [29] where vehicles are assumed to not have the full required packages needed for the requested tasks.

In this paper, we extend our previous work in [25], where the focus was on minimizing the total power consumption, to consider joint minimization of the total power consumption, propagation delay, and queuing delay in a cloud-fog-VEC architecture. In [30], we presented the idea of the proposed joint optimization model and discussed the preliminary results that are also extended in this paper.

Previous research developed a joint optimization model with power and delay as part of the multi objective optimization. For example, the authors in [31] proposed a fitness estimation model to allocate tasks to a network of three layers, which include the central cloud, cloudlet nodes, and VEC. This model was designed using three objectives to minimize network delay, minimize power consumption, and maximize the availability of the processing resources (as virtual machines). The total network delay is calculated using the propagation, queuing, transmission, and processing delay. The queuing delay considered is based on the waiting time of the processing tasks in each processor and is calculated based on the total execution time. This is the same for the power consumption, which is calculated only based on the power needed to execute each allocated task. The study mainly focused on the proposed fitness estimation model, which studied the capability of the available resources and their availability, a problem that is considered a challenge due to the vehicles' mobility.

The work in [32] proposed a joint optimization model for task allocation, to minimize service delay and quality loss. The optimization model considered the presence and absence of a fog node and its capacity within the studied area. They minimized the maximum service delay which is composed of the time needed for transmission and processing, considering the maximum tolerable delay for video streaming applications. The authors investigated the trade-off between the service delay and the computing quality loss ratio and achieved a balanced optimization between the two objectives. Due to the model complexity in testing real time traffic, they based their evaluation on an optimization-based heuristic algorithm considering two periods with different vehicle densities. Their results found that the developed algorithm reduced the service delay by up to 27% and enhanced the quality loss ratio by 56%. The study in [33] proposed a reliable task scheduling model in the vehicular cloud in order to minimize the execution time and satisfy job deadlines using mixed integer linear programming (MILP). They considered the MapReduce execution model to distribute big data processing jobs among the available vehicles without considering any fixed node. This allocation was made by the VEC cluster head, which is a selective vehicle located around the participant vehicles. As these vehicles are in a dynamic mode, the quality of the provided service depends on how reliable the connection is with the cluster head, and how much time is needed to process and send the output back to this cluster head. These two factors were investigated by minimizing the response delay and maximizing the success probability. The optimized model was also developed using a simulation-based algorithm to test the solution, one that considered vehicle density and an increase in the level of collision in order to study the system's reliability. It was found that the increased number of vehicles reduced the job execution time and increased the percentage of successful job execution. In conclusion, the proposed scheduling solution achieved a faster time and satisfied the required level of reliability.

In another study in [34], the authors proposed an efficient task-scheduling model in the vehicular cloud to minimize the average completion time. In the proposed VEC scenario, a fixed fog node offloaded a job (divided into many tasks) to the available surrounding VEC cluster. The tasks are offloaded simultaneously and processed in parallel, where each vehicle executes a task, sends the generated output to the vehicle executing the next task, and so on. The minimized delay in the proposed model consists of processing time, communication time and queuing time. The latter time is calculated based on the generation time of the first task, the processing time, and the completion time of the last task. The evaluated results are based on a comparison between the optimization-based algorithm developed and the previous algorithm from their literature (greedy algorithm). It is found that their developed algorithm can reduce the average response time by 12% compared to the greedy algorithm.

The remainder of this paper is organized as follows: In Section II, we present the cloud-fog-VEC architecture and define the delay aspects of the network which are included in the joint optimization. In Section III, the joint optimization model is described. The scenarios considered and the joint optimization results are presented and discussed in Section IV. Finally, Section V concludes the paper and highlights some of the planned future work.

## II. THE PROPOSED CLOUD-FOG-VEHICULAR EDGE CLOUD ARCHITECTURE

The cloud-fog-VEC architecture considered is shown in Figure 2. It is composed of four distinct layers with four processing locations at the core, metro, access, and edge layers. We consider the architecture with processing resources at the central cloud (CC), the metro fog (MF), the OLT fog (LF), the ONU fog (NF) and the vehicular edge cloud (VEC). The edge layer consists of one VEC cluster and multiple source nodes (SNs)

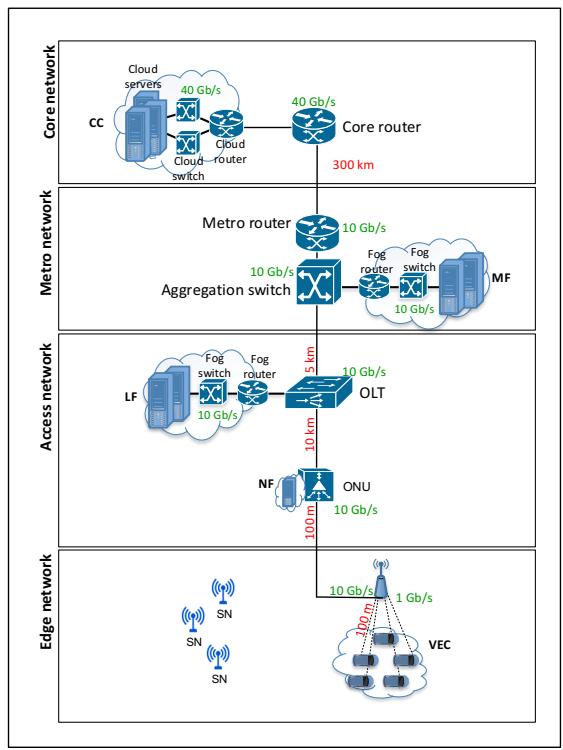

**Figure 1** End-to-End Cloud-Fog-VEC Architecture, with estimated distances (in red) and network devices service rate (in green).

We study the propagation and queuing delay. The distances illustrated in Figure 1 (colored in red) are used to estimate the propagation delay. The distances are based on the following assumptions:
1. The distance between the AP and surrounding VNs is set to the standard coverage range of IEEE 802.11, namely 100 m [35].
2. The distance between the AP and ONU is estimated based on the observation that one ONU can connect to multiple

APs in a wired local area network (LAN) with a maximum of 100 m distance [36].
3. The distance between the ONU and OLT is based on typical PON designs in the field [37]. Here, we considered a design where the OLT is located in the telecom main office in the centre of the city. ONUs usually represent devices located at the end-users' location (i.e., at home); usually such distance are around 5–20 km [37] so we assumed a distance equal to 10 km.
4. The distance between the OLT and metro node (router and switch) was estimated based on the metro network design. The metro network usually has a radius of 20–120 km [38]. The OLT can be either collocated with the metro node in the same telecom office or located somewhere else in the local area of the metro node (1–10 km away). We based our estimation on the latter scenario with an approximate distance equal to 5 km between the OLT and metro node. The MF node here is assumed to be located within few kilometres of the metro node. The same distance is assumed to exist between OLT and LF.
5. The distance between the metro node and the core node (including the associated CC), is given as the distance between two large cities, assuming the current city does not have a large central cloud. An example of such a distance is taken as the distance between Leeds and a large data centre in London, a 300 km distance.

The queuing delay was modelled for each networking node as an M/M/1 queue with one server, where arrivals follow a Poisson process and the service rate is negative exponentially distributed, described in Figure 2.

The propagation delay was calculated for each network location. The queueing delay was calculated based on traffic and the capacity of the nodes. Here the aggregated traffic delivered, and the node or device maximum service rate defined the arrival rate and service rate, respectively, and are used to determine the queueing delay as given below

$$Queuing\ Delay = \frac{1}{\mu - \lambda} \quad (1)$$

where $\mu$ is the service rate, and $\lambda$ is the arrival rate (summation of the traffic delivered assuming a Poisson process) in the network device. We have considered in this work delay at the packet level. We used the Ethernet maximum packet size of 1500 bytes and therefore expressed the arrival data rates in packets per second and expressed the service rates (transmission rates) in packets per second.

It is worth noting that, based on our proposed architecture, the AP should work as a network coordinator to connect the edge layer to the wired infrastructure, and therefore, deliver the processing tasks to the optimum PN. Accordingly, we assumed that the AP works at two different service rates based on the interface used; either the wired fiber infrastructure (with a 10 Gb/s service rate), or wireless medium (with 1 Gb/s service rate). Moreover, the core node, with the associated data centre, was assumed to work at 40 Gb/s, as they are part of the IP/WDM network. Other network devices were assumed to have a 10 Gb/s service rate based on GPON. The services rate values are shown in Figure 1, (colored in green).

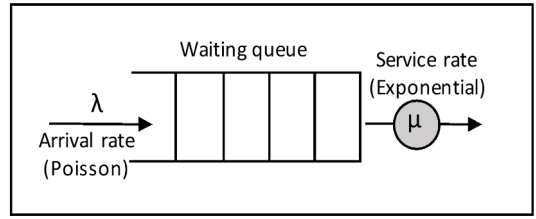

**Figure 2** M/M/1 Queueing model.

### III. MILP MODEL

The MILP model introduced in [25] was extended to jointly minimize power consumption and delay. To continue to use linear programming, Equation (1) was converted to a linear form using a lookup table [39]. The lookup table is predefined with all the possible generated traffic combinations (arrival rates indicator), indexed with the calculated queuing delay based on a fixed service rate. As we have three different service rate values in our designed network, three lookup tables were defined. Based on this arrival rate indicator, the queuing delay for a node was given as the value corresponding to the indicator in the lookup table.

The modified MILP defines the following additional sets, parameters, and variables:

The sets, parameters, and variables are declared as follows:

| Set: | |
|---|---|
| $A$ | Set of AP nodes considering the wired interface, where $A \subset N$. |
| $AL$ | Set of AP nodes considering the wireless interface, where $AL \subset N$. |
| $AR$ | Set of arrival rates. |
| $SR$ | Set of service rates. |
| **Parameters:** | |
| $H_{as}$ | Queuing delay at arrival rate $a \in AR$ and service rate $\in SR$, in the lookup table. |
| $G1$ | Large enough number with units of Mb/s. |
| $G2$ | Large enough number with units of ms. |
| $D_{ij}$ | Distance between any two nodes $(i,j)$, where $i \in N$, $j \in NN_i$. |
| $\mathbb{C}$ | Speed of light, $\mathbb{C} = 299{,}792 \frac{km}{s}$. |
| $\Delta RI$ | Refractive index of fibre, it defines the ratio of the speed of light in fibre to speed of light in free space; $\Delta RI = \frac{2}{3}$. |
| **Variables:** | |
| $\zeta_{ij}^{sd}$ | Binary variable $\zeta_{ij}^{sd} = 1$ if traffic flow sent from source node $s$ to processing node $d$ traverses physical link $(i,j)$, where $s \in SN$, $d \in PN$, and $i,j \in N$. |
| $Q_{ij}^{sd}$ | Queuing delay at node $j$ experienced by the traffic from source node $s$ to processing node $d$ traversing physical link $(i,j)$, where $s \in SN$, $d \in PN$ and $i,j \in N$. |
| $Q_i$ | Queuing delay experienced by traffic aggregated at |
| | node $i \in N$. |
| $Q_{sd}$ | Queuing delay of the traffic sent from source node $s \in SN$ to processing node $d \in PN$. |
| $Q$ | Total queuing delay of the network. |
| $R_{sd}$ | Propagation delay of the traffic sent from source node $s \in SN$ to processing node $d \in PN$. |
| $R$ | Total propagation delay of the network. |
| $\lambda_i$ | Arrival rate (total traffic) at each node $i \in N$. |
| $\sigma_{ij}$ | Arrival rate indicator for node $i \in N$, $\sigma_{ij} = 1$ if the arrival rate of node $i$ matches rate $j \in AR$, it is 0 otherwise. |

All the power consumption equations in [25], ie equations (2) to (21), were considered in this model. The total power which results from adding all these power consumption components is as follows:

$$P = TPC^{CC} + TPC^{MF} + TPC^{LF} + TPC^{NF} + TPC^{VN} + TPC^{NET} \quad (2)$$

Additionally, the following equations are used to calculate the propagation and queuing delay for the network.

**1) The total propagation delay (R),** is calculated based on the propagation delay between all source node and processing node pairs and is given as

$$R = \sum_{s \in SN} \sum_{d \in PN} R_{sd} \quad \forall \ s \in SN, d \in PN \quad (3)$$

where $R_{sd}$ is is the propagation delay of the path traversed by traffic sent from each source node $s \in SN$ to the processing node $d \in PN$, and is calculated as follows;

$$R_{sd} = \sum_{\substack{i \in N \\ i \notin AL}} \sum_{\substack{j \in Nmi \\ j \notin VN}} \zeta_{ij}^{sd} \frac{D_{ij}}{\Delta RI \ \mathbb{C}} \quad (4)$$

$$\forall \ s \in SN, d \in PN, d \notin VN$$

$$R_{sd} = \sum_{\substack{i \in N \\ i \in AL}} \sum_{\substack{j \in Nmi \\ j \in VN}} \zeta_{ij}^{sd} \frac{D_{ij}}{\mathbb{C}} \quad \forall \ s \in SN, d \in VN. \quad (5)$$

Equation (4) and (5) calculate the propagation delay for the traffic sent to the processing nodes via fibre or wireless links, respectively. A refractive index $\Delta RI$ with the value of $\frac{2}{3}$ is added to Equation (4) to define the ratio of the speed of light in fibre to the speed of light in free space.

**2) The total queuing delay (Q),** which is calculated based on the queuing delay experienced by traffic between all the source node and processing node pairs, and is given as

$$Q = \sum_{s \in SN} \sum_{d \in PN} Q_{sd} \quad \forall \ s \in SN, d \in PN \quad (6)$$

where $Q_{sd}$ is the queuing delay of the path traversed by traffic sent from each source node $s \in SN$ to processing node $d \in PN$, and is calculated as

$$Q_{sd} = \sum_{\substack{i \in N \\ i \notin SN \cup AL}} \sum_{j \in Nmi} Q_{ij}^{sd} \quad \forall \ s \in SN, d \in PN, d \notin VN \quad (7)$$

$$Q_{sd} = \sum_{\substack{i \in N \\ i \notin SN \cup A}} \sum_{j \in Nmi} Q_{ij}^{sd} \quad \forall \ s \in SN, d \in VN. \quad (8)$$

Equations (7) and (8) calculate the queuing delay for a traffic demand by summing the queuing delay experinced by the demand at each node. As mentioned earlier, the traffic handled by the AP can be sent via one of two interfaces; wired interface at 10 Gb/s and wirless interface at 1 Gb/s; thus, equation (7) considers the traffic sent via the wired interface of the AP while equation (8) considers the traffic sent to the vehicular nodes via the 1G wireless interface.

The joint objective is defined as
**Minimize**

$$\alpha P + \beta R + \gamma Q \quad (9)$$

where $\alpha$, $\beta$, and $\gamma$ are weight factors used for the following purposes: (i) to scale the terms so that they are comparable in magnitude; (ii) to emphasize and de-emphasize terms (power, queuing delay and propagation delay); and (iii) to accommodate the units in the objective function. Therefore, $\alpha$ is a unitless factor, and $\beta$ & $\gamma$ have units of $\frac{Watt}{sec}$.

In addition to the constraints in [25], the model is subject to the following additional constraints:

1) The traffic estimation at each node

$$\sum_{s \in SN} \sum_{d \in PN} \sum_{j \in Nmi} \lambda_{ij}^{sd} = \lambda_i \quad \forall \ i \in N, i \notin A \cup AL \quad (10)$$

$$\sum_{s \in SN} \sum_{d \in PN} \sum_{j \in Nmi \cap U} \lambda_{ij}^{sd} = \lambda_i \quad \forall \ i \in A \quad (11)$$

$$\sum_{s \in SN} \sum_{d \in PN} \sum_{j \in Nmi \cap VN} \lambda_{ij}^{sd} = \lambda_i \quad \forall \ i \in AL \quad (12)$$

Constraint (10) calculates the traffic arrival at each node in the network except APs. Constraints (11) and (12) estimate the arrival traffic at wired and wireless AP interfaces, respectively.

2) The arrival rate indicator:

$$\sum_{j \in AR} H_{ij} \ j = \lambda_i \quad \forall \ i \in N, i \notin A \cup AL \quad (13)$$

$$\sum_{j \in AR} H_{ij} \ j = \lambda_i \quad \forall \ i \in A \quad (14)$$

$$\sum_{j \in AR} H_{ij} \ j = \lambda_i \quad \forall \ i \in AL \quad (15)$$

Constraints (13) to (15) create indicators of the arrival rate for each node. This is equal to 1 if the arrival rate is equal to $j$:

$$\sum_{j \in AR} H_{ij} \leq 1 \quad \forall \ i \in N, i \notin A \cup AL \quad (16)$$

$$\sum_{j \in AR} H_{ij} \leq 1 \quad \forall \ i \in A \quad (17)$$

$$\sum_{j \in AR} H_{ij} \leq 1 \quad \forall \ i \in AL \quad (18)$$

Constraints (16) to (18) ensure that each node has no more than one arrival rate indicator for a given service rate.

3) Queuing delay estimation:

$$\sum_{j \in AR} H_{ij} \cdot \eta_{js} = Q_i \quad (19)$$
$$\forall\ i \in RR \cup CR \cup CS,\ s = \frac{40Gb}{s}$$

$$\sum_{j \in AR} H_{ij} \cdot \eta_{js} = Q_i \quad (20)$$
$$\forall\ i \in N, i \notin RR \cup CR \cup CS \cup A \cup AL,\ s = \frac{10Gb}{s}$$

$$\sum_{j \in AR} H_{ij} \cdot \eta_{js} = Q_i \quad \forall\ i \in A,\ s = \frac{10Gb}{s} \quad (21)$$

$$\sum_{j \in AR} H_{ij} \cdot \eta_{js} = Q_i \quad \forall\ i \in AL, s = \frac{1Gb}{s} \quad (22)$$

Constraints (19) and (20) estimate the traffic delay for each node that operates at 40 Gb/s or 10 Gb/s, respectively. Constraints (21) and (22) estimate the delay for the AP wired and wireless interfaces, respectively.

$$\lambda_{ij}^{sd} \geq \zeta_{ij}^{sd} \quad \forall\ s \in SN, n \in PN \quad (23)$$

$$\lambda_{ij}^{sd} \leq G1\ \zeta_{ij}^{sd} \quad \forall\ s \in SN, n \in PN \quad (24)$$

Constraints (23) and (24) set $\zeta_{ij}^{sd} = 1$ if the traffic demand between the source node $s \in SN$ and the processing node $d \in PN$ is routed through link $(i,j) \in N$

$$Q_{ij}^{sd} = Q_i\ \zeta_{ij}^{sd} \quad \forall\ s \in SN, d \in PN \quad (26)$$

$$Q_{ij}^{sd} \leq G2\ \zeta_{ij}^{sd} \quad \forall\ s \in SN, d \in PN \quad (27)$$

$$Q_{ij}^{sd} \leq Q_i \quad \forall\ s \in SN, d \in PN \quad (28)$$

$$Q_{ij}^{sd} \geq Q_i - G2\left(1 - \zeta_{ij}^{sd}\right) \quad \forall\ s \in SN, d \in PN \quad (29)$$

Equation (25) calculates the queuing delay at node $j$ for the traffic sent from source node $s$ to processing node $d$, through node $i$. As equation (25) involves the multiplication of two variables, $Q_{ij}^{sd}$ and $Q_i$, constraints (26) to (28) are used to remove the nonlinearity of equation (25) and replace the relationship with an equivalent linear relationship.

## IV. EVALUATION SCENARIOS AND RESULTS

### A. Scenarios considered

The model presented in the previous section is considered with the following variations of the objective function:
1) Minimizing the total power consumption only, by setting $\alpha$ to 1; and $\beta$ and $\gamma$ to zero in equation (9).
2) Minimizing the traffic propagation delay only, by setting the values of $\beta$ to 1; and $\alpha$ and $\gamma$ to zero in equation (9).
3) Minimizing the power consumption and traffic propagation delay jointly, by setting the values of $\alpha$ to 1, $\gamma$ to zero, and $\beta$ to a value that ensures equal importance of the power consumption and propagation delay. This was done as follows:
  a) Running the MILP model with the minimizing power consumption objective.
  b) Running the MILP model with the minimizing propagation delay objective.
  c) Calculating $\beta = \frac{P}{R}$ such that $\alpha P = \beta R$, where $P$ is the power consumption and $R$ is the propagation delay.
  d) Runing the MILP model with the objective of minimizing both the power consumption and propagation delay and comparing the $\alpha P$ and $\beta R$ values.
  e) Adjusting the value of $\beta$ again, so that the joint objective function produced the required equality ie $\alpha P = \beta R$.

4) Minimizing the traffic queuing delay, by setting the values of $\alpha$ and $\beta$ in equation (9) to zero, and the value of $\gamma$ to 1.
5) Minimizing the power consumption and traffic queuing delay jointly, by setting the values of $\alpha$ to 1 and $\beta$ to zero, and $\gamma$ to a value that ensures equal importance of the power consumption and queuing delay. This was done by repeating the same steps mentioned previously in 3), replacing $\beta$ and $R$ with $\gamma$ and $Q$, respectively. This was carried out to ensure that the following equality $\alpha P = \gamma Q$ was achieved.
6) Minimizing the power consumption, traffic propagation and queuing delay jointly, by setting the values of $\alpha$ to 1, and the values of $\beta$ and $\gamma$ to two ratios, to ensure that $\alpha P = \beta R = \gamma Q$, using, again, the same steps described in 3).

The previously described objective functions were combined into four scenarios that highlight the individual effects of the propagation and queuing delay, combined with the power consumption on the processing allocation decision, and both power and delay values. All the scenarios considered a cloud-fog-VEC allocation (CFVA) with low-density VNs (8 VNs) and single allocation (no task splitting). We assessed the allocation problem with ten generated tasks having low processing demands (100–1000 MIPS) and a fixed data rate ratio (DRR) of 0.1. The reason for choosing a low demand with a high DRR was to generate intensive traffic (10–100 Mb/s) per task, in order to study the delay results in such a congested network. It is worth mentioning that the queuing delay value is expressed in sec/packet as given in (1), where we used the Ethernet MTU packet size of 1500 Byte.

**SCENARIO#1. Power and Propagation Delay Minimization**

In this scenario, we study the joint minimization of the power consumption and the propagation delay (objective function case 3), and compare the results to the two cases where only the power (objective function case 1) or propagation delay (objective function case 2) are minimized. Figures 3 and 4 illustrate the total power consumption and the average propagation delay for the three cases considered in the objective function versus the total traffic generated from the ten tasks, each 100–1000 Mb/s.

Figure 3 shows that the power minimized case produced the lowest power consumption. The jumps in the curves in the three minimization functions are due to moving the allocation at higher traffic to a less efficient PN that can support the traffic.

For example, in the power minimized case, the small jumps at 400 Mb/s were caused by moving the allocation from VN to NF, as seen in Figure 5. The optimization did this because of the limited VN connection data rate (72.2 Mb/s per VN) which cannot serve more than one task with the 40 Mb/s required data rate. Activating the ONU and its processor (NF) caused all the tasks to be allocated to the NF instead of activating two locations. Another power rise occurred at 800 Mb/s, when all tasks were allocated to the LF. The delay minimized objective led to higher power consumption (by up to 35%) compared to the power minimized case (Figure 3). This is due to all (or the majority of) tasks being allocated in VNs, even with the NF being activated (Figure 5), which is not a power-efficient allocation decision. With the joint minimization of the power and delay, the MILP results (at 100–700 Mb/s) led to a power consumption comparable to the power minimized optimization, due to the fact that minimizing delay required poor energy-efficiency choices, i.e., the placement of tasks in two PN (NF & VN), which was not allowed by a MILP that weighs power and delay equally. However, the power consumption results beyond 700 Mb/s became comparable with the delay minimized optimization, as the propagation delay became a limiting factor which causes the model to allocate tasks to both locations (NF and LF) in order to achieve a lower average delay over all of the tasks. It is worth mentioning that the power consumption resulting from the joint optimization (the green curve in Figure 3) is not exactly in the middle of the power minimized and delay minimized curves, as might be expected, because the number of placement options (PNs) was finite and discrete. Practical systems will typically have a similar number of processing locations (PNs) or even fewer as building fog processing nodes at a higher granularity (e.g. few hundred meters) is not practical.

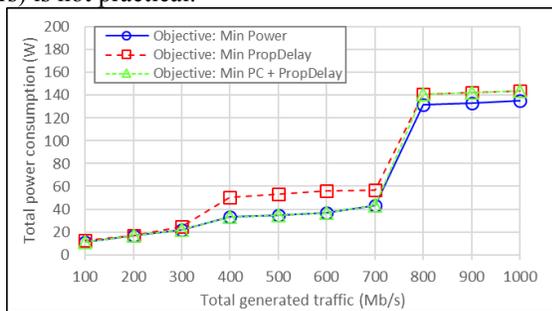

**Figure 3** Total power consumption (Scenario 1)

Figure 4 displays the average propagation delay for the same three objective function cases. For the power minimized case, it can be seen that the low power consumption in Figure 3 has resulted in the high level of propagation delay in Figure 4. The delay starts low (similar to the other two optimization cases), as the tasks are allocated in the three cases to VNs, as VNs are the most efficient PN for both power and propagation delay minimization. However, the delay increases with increase in the traffic as the tasks are allocated to further locations (i.e., NF and LF). It can also be seen from the zoomed-in figure (in Figure 4) that tasks allocated to NF (at 300–700 Mb/s) experienced more propagation delay compared to the VN allocation, despite the fact that both the PNs were 100 m away from the AP (as mentioned in Section 2). This disparity is attributed to the fact that the speed of light in fibre, connecting the AP to NF, is affected by a refractive index of 2/3, as defined in Equation (4), which causes an increase in the propagation delay of the NF compared to VN. Thereafter, the delay jumps from 0.5 µs to 50 µs when the allocation changed from NF to LF (beyond 700 Mb/s), as the LF is located 10 km away from the ONU. This shows the effect of the PN location on the propagation delay and therefore, on the allocation decision. This effect was confirmed when minimizing the propagation delay, as seen in Figure 4, when the model allocated more tasks to the available VNs even if other PNs nodes were activated (as seen in Figure 5). Similar to the power results shown in Figure 3, minimizing both the power and delay showed an average delay relatively comparable to the case of minimized power at 100–700 Mb/s, and to the case of minimized delay beyond 700 Mb/s. This is due to the same reason mentioned previously, as the finite number of available processing locations causes comparable results to either one of the other cases rather than having results in between the two other cases.

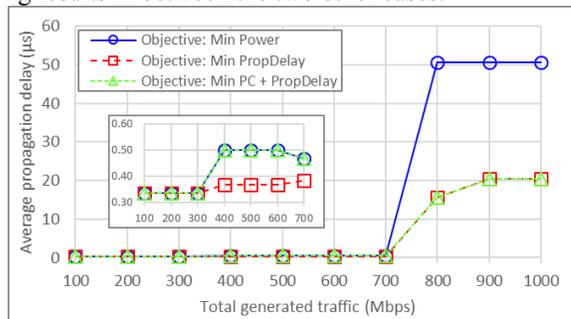

**Figure 4** Average propagation delay (Scenario 1), enclosed zoom-in figure shows the result in 100–700 Mb/s range.

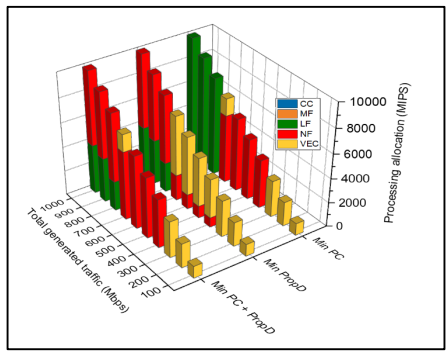

**Figure 5** Processing allocation in each PN (Scenario 1)

**SCENARIO#2**. **Power and Queuing Delay Minimization.**
In this scenario, we study the joint minimization of the power consumption and the queuing delay (objective function case 5), and compare it to the power minimized case (objective function case 1), and the queuing delay minimized objective function (objective function case 4). Figure 6 shows the total power consumption for the three cases. Relatively comparable power consumption results can be observed for the three optimization objectives. The power consumption minimized case shows an increase in the power consumption whenever the allocation is changed to a less efficient location (similar to the previous result shown in Figure 3). When minimizing the queuing delay, the results showed an early increase in the total power consumption as the VNs were avoided and the tasks were allocated to the NF, as seen in Figure 8. This allocation decision is attributed to the bad service rate of the AP wireless interface, which causes a very high queuing delay at this interface and therefore, the VN becomes an inefficient location in terms of the average queuing delay. This allocation explains the comparable power consumption, in this case, to the power minimized case (at 400–600 Mb/s), as the tasks in both cases were allocated to NF. However, as the generated traffic increases (beyond 600 Mb/s), the power consumption jumps to 140 W, an increase by 70%, compared to the power minimized case. At this point, the NF became exhausted and was unable to accommodate all the generated tasks, and thus, the LF was activated to accommodate the remaining workload. This explains the continued and saturated increase in the power consumption (by 7%) compared to power minimized case (at 800–1000 Mb/s). In the latter case, activating one PN to serve all the generated tasks is more efficient than activating two or more PNs. This is because activating each fixed fog node consumes extra power overhead due to the idle power and PUE of the network devices and servers. In the case of the joint minimization of both the power consumption and queuing delay, the power consumption results were very comparable to the queuing delay minimized case. Including the queuing delay in the objective function avoids allocating any tasks to VN, as it causes a significant increase in the delay. One case resulted in the allocation of a portion of the processing workload to VNs, at 700 Mb/s, where the model achieved a balance between the power and the delay by allocating a small portion of the workload to VNs (140 Mb/s) with the majority of the workload allocated to NF (560 Mb/s). This occurred when the NF became exhausted (with 5600 MIPS allocation) and was not able to bin-pack any more workload to achieve full utilization; hence, the optimization allocated the remaining 1400 MIPS to VN, as activating the LF would cause a significant increase in the power consumption and therefore, the objective function balance might not be achieved.

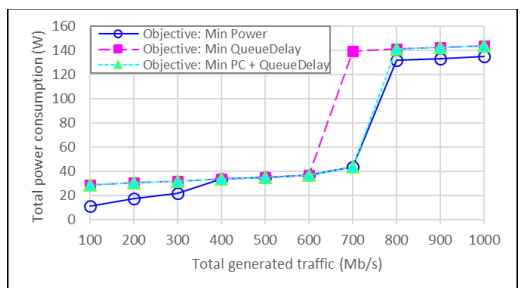

**Figure 6** Total power consumption (Scenario 2)

Figure 7 shows the average queuing delay for the same three objective function cases. As described earlier in Section 1, the queuing delay of a node is influenced by the service rate of this node, hence, the PNs with a high service rate (central cloud and fixed fog nodes) had a lower delay, compared to the wireless-based VNs, due to the low service rate of the wireless devices (i.e., AP). For example, in the power minimized case (in Figure 7), low demand tasks were allocated to VNs (as seen in Figure 8). This caused a high average queuing delay (13–17 μs) due to the low AP service rate (1 Gb/s). As soon as the traffic increased and the tasks were allocated to NF (as VN became insufficient), the average queuing delay dropped by 85% (from 17 μs to 2.5 μs) and saturated at this value, which holds as long as the NF remains the most power efficient placement. The queuing delay experienced a continued increase due to the increase in the number of hops to the optimum allocation at the LF. With the queuing delay minimized objective, the delay was reduced by an average of 84%, compared to the power minimized case at low generated traffic. This reduction was due to the allocation of tasks to the fixed fog (NF), with a better service rate (10 Gb/s). The model maintained a continuous low average queuing delay with increase in the total traffic, which became comparable with the power minimized case, and then increased slightly due to part of the task being allocated to an extra PN. However, the model yielded up to 43% lower average queuing delay with high traffic, compared to the power minimization case. The case where both the power and delay were minimized produced a comparable average delay to the delay minimized case. As mentioned previously with the power consumption results, the queuing delay became a limiting factor that constrained the optimization and resulted in the tasks being allocated to VNs, due to the bad service rate of the AP wireless interface, which caused a huge increase in the queuing delay. One exception occurred at 700 Mb/s, where the balance was achieved by allocating the tasks to two PNs (i.e. NF and VN) due to NF insufficient capacity, as seen in Figure 8.

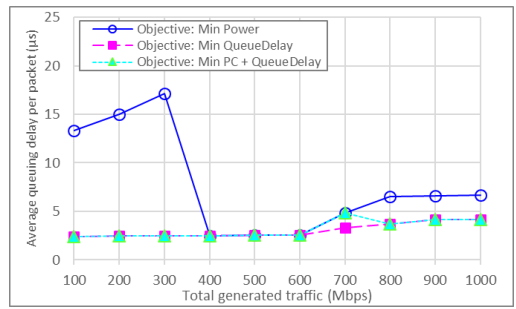

**Figure 7** Average queuing delay (Scenario 2)

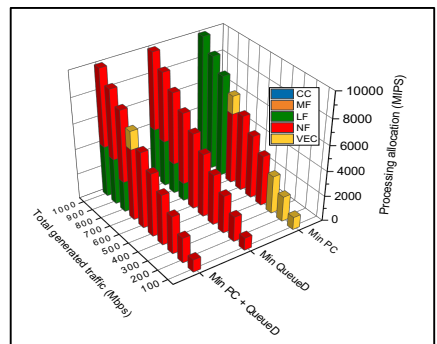

**Figure 8** Processing allocation in each PN (Scenario 2)

### SCENARIO 3. Queuing Delay Optimization with Multiple AP Service Rate

As observed previously in Scenario 2, the queuing delay in the access layer became a limiting factor which made the allocation of tasks to the VNs unfeasible due to the wireless devices' insufficient service rate. The AP is the control gate for VNs, it is thus important to investigate the effect of the multiple service rates on the VNs allocation and the average queuing delay. Two objective functions were examined: minimizing the queuing delay (Figures 9–11) and minimizing the power and queuing delay, jointly (Figures 12–14). In each optimization function setting, three cases were investigated where the AP wireless interface was set to 1, 5 and 10 Gb/s service rate. The case where the AP operated at 1 Gb/s was the default service rate, and its related results were discussed earlier in Scenario 2 (Figures 6–7) and are provided in this scenario as a baseline case. The case where the AP operated at 10 Gb/s represented the other extreme where the AP has a high service rate equivalent to the service rate of the wired interface. The third case where the AP service rate was 5 Gb/s provided an average service rate between the other two cases. This helps in the determination of the impact of the service rate, and in particular if a higher service rate is not required to achieve a lower queuing delay.

Figures 9 and 10 show the total power consumption and average queuing delay when the queuing delay was minimized at the three defined service rates for the AP wireless interface. The results show that with an increased service rate, the queuing delay is reduced and the VN allocation increased. This may increase or decrease the power consumption based on the number of activated PNs. For example, at 5 Gb/s service rate, the restriction in the VNs allocation was relaxed, and the tasks were allocated to the VN alongside the NF (as in Figure 11), which justifies the small increase in power consumption (an average of 14%), compared to the 1 Gb/s case. However, a power consumption saving of 67% was observed at 5 Gb/s service rate at a 700 Mb/s incoming traffic rate, as some tasks were allocated to VN instead of LF in the 1 Gb/s case, at the same 700 Mb/s traffic. On the other hand, the average queuing delay in the 5 Gb/s service rate case (in Figure 10) did not achieve a significant improvement up to an incoming traffic rate of 600 Mb/s despite the increase in the AP service rate. This is attributed to the accumulated queuing delay based on the number of hops in the route leading to the PN.

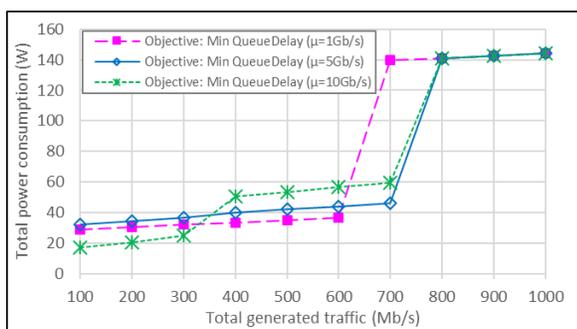

**Figure 9** Total power consumption, under average queuing delay minimization objective, with multiple AP service rates (Scenario 3).

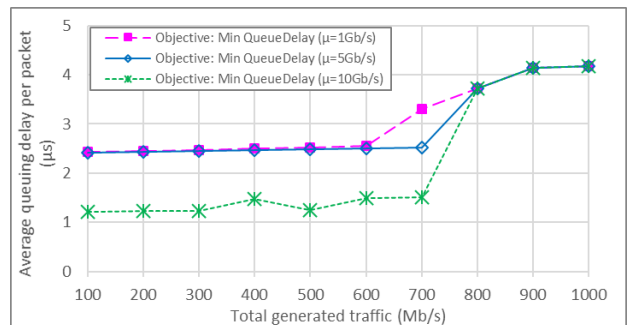

**Figure 10** Average queuing delay, under average queuing delay minimization objective, with multiple AP service rates (Scenario 3).

For example, at 300 Mb/s traffic and with 1 Gb/s service rate, all ten tasks were allocated to the NF (as seen in Figure 11). The resulting average queuing delay was 2.5 µs, caused by the delay of the route leading to the NF (including the AP wired interface and the ONU). On the other hand, at the same 300 Mb/s traffic and with 5 Gb/s service rate, the tasks were allocated to the two PNs (LF and VN). The average queuing delay over all the tasks was equal to 2.4 µs, with only 0.1 µs improvement. This limited reduction in the queuing delay was due to a comparable queuing delay resulting from the two allocations because the decrease achieved by the better AP service rate resulted in a comparable queuing delay of the two hop route to the NF (AP and ONU). By increasing the AP wireless service rate to 10 Gb/s, we saw an early reduction in the power consumption (Figure 9) as all the tasks were fully allocated to VN (as shown in Figure 11). However, the power consumption increased afterwards because of two reasons: Firstly, due to the activation of two PNs located at the NF and VN, and secondly due to the utilization of more VNs, which caused extra power overhead resulting from activating the VN communication adapter. This resulted in the increase in power consumption by 34% and 21%, compared to the cases with 1 and 5 Gb/s service rates. However, the average queuing delay (Figure 10) shows significant reduction by an average of 48% compared to the 1 Gb/s case, because the service rate of both the AP interfaces became equivalent. Hence, the VN which is one hop away from the AP became more efficient than the PN that is two hops away (i.e., the NF).

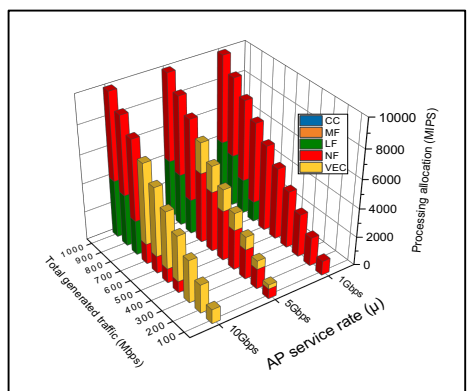

**Figure 11** Processing allocation in each PN, under average queuing delay minimization objective, with multiple AP service rates (Scenario 3).

Figures 12 and 13 show the total power consumption and the average queuing delay when the power and queuing delay were jointly minimized with the same three AP's service rates. The results show the small effect of the AP service rate on power consumption. This effect was only observed at low generated traffic, as the increase in the AP service rate relaxed the restriction of allocating tasks to VNs. Therefore, these VNs became an optimum location for both the queuing delay and power consumption. Moreover, a reduction in the average queuing delay (in Figure 13) was achieved whenever more tasks were allocated to the VNs with a high AP service rate. This reduction became significant when the tasks were allocated to the VN alongside the NS. This can be seen at 700 Mb/s traffic, where the average queuing delay was 4.8 µs when allocating tasks to VNs (as shown in Figure 14) due to the poor AP service rate. Although the optimization resulted in the same allocation decision, with 5 Gb/s and 10 Gb/s service rates, the average queuing delay dropped by 47% and 52%, respectively, compared to the 1 Gb/s service rate case. This confirmed that the increase in the AP service rate had a small effect on the allocation decision in the power and delay joint optimization.

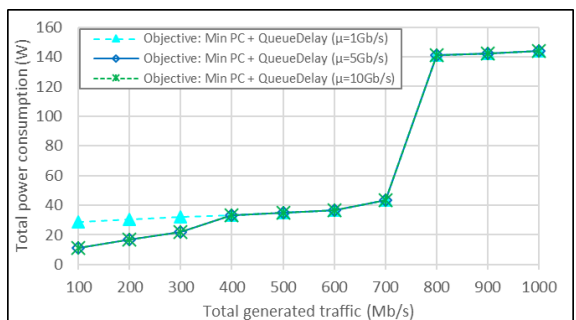

**Figure 12** Total power consumption, under power consumption and average queueing delay minimization objective, with multiple AP service rates (Scenario 3).

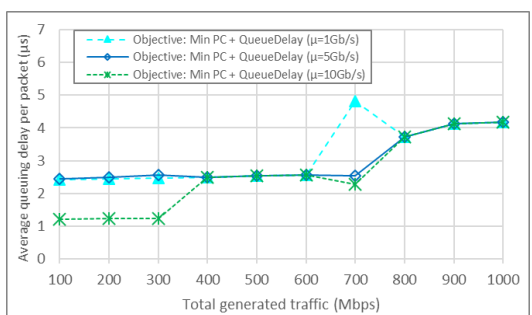

**Figure 13** Average queuing delay, under power consumption and average queueing delay minimization objective, with multiple AP service rates (Scenario 3).

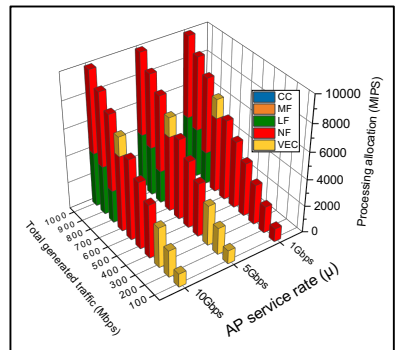

**Figure 14** Processing allocation in each PN, under power consumption and average queueing delay minimization objective, with multiple AP service rates (Scenario 3).

**SCENARIO 4. Power, Propagation Delay and Queuing Delay Multi-Objective minimization**

In this scenario, we study the joint optimization of the power consumption, propagation delay and queuing delay, compared to the joint optimization of the power consumption and propagation delay discussed in Scenario 1 and the joint optimization of the power consumption and queuing delay discussed in Scenario 2

Figure 15 shows the total power consumption for the three objective function cases considered. The results show relatively comparable power results for the three objectives. We noticed, however, that whenever the optimization objective included the queuing delay, an immediate restriction was applied on allocating any workload to the VNs, as shown in Figure 18. This is attributed to the fact that the allocation in VN causes a very high queuing delay at the AP wireless interfaces, which excludes the VNs from the allocation decision. Therefore, the multi-objective function optimization followed the same outcomes of the power minimized optimization and the queuing delay objective when considering the average propagation delay and the average queuing delay results (Figures 16 and 17).

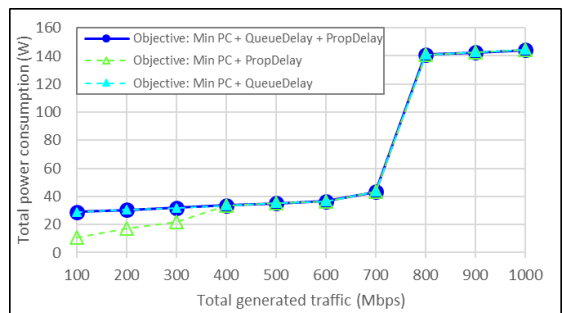

**Figure 15** Total power consumption (Scenario 4)

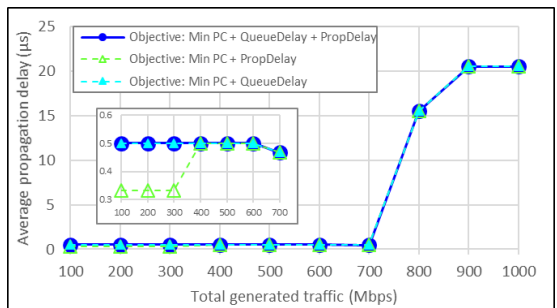

**Figure 16** Average propagation delay (Scenario 4)

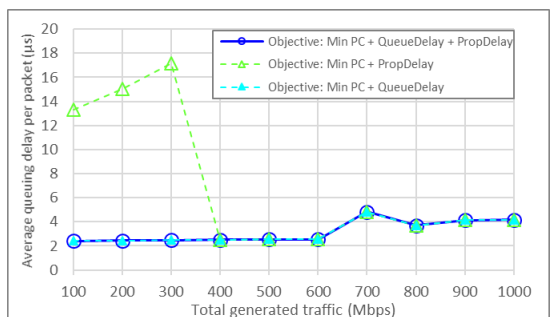

**Figure 17** Average queuing delay (Scenario 4)

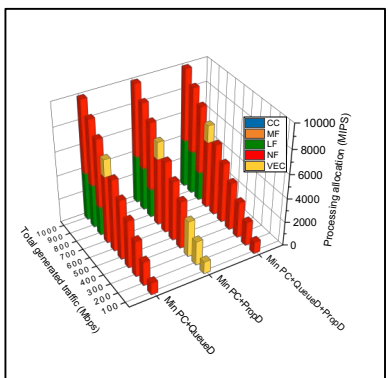

**Figure 18** Processing allocation in each PN (Scenario 4)

## V. CONCLUSIONS

In this paper, we have investigated the joint optimization of power consumption, propagation delay and queueing delay when allocating processing tasks to cloud, fog and VEC processing nodes in an architecture which provides multiple processing locations. The study was carried out by extending the MILP introduced in [25], where the propagation delay and queuing delay are added. The evaluation considered different cases of the objective function where power consumption, propagation delay and queuing delay are examined separately or together.

Our results show that the closer the PN is to the AP, the lower the power consumption and delay, as the distance and number of hops affect the propagation delay and queuing delay. However, the queuing delay at the AP becomes a limiting factor when it operates at a low service rate compared to the traffic arrival rate. Thus, the allocation of processing tasks at the VN was avoided whenever the optimization objective function included queueing delay and the AP operated at a low service rate. Increase in the AP service rate results in a lower queuing delay and better VN utilization.

Future work can introduce additional optimization components to the delay such as processing and transmission (ie packet serialization) delay alongside propagation and queuing delay. It can also consider more detailed AP queueing models such as models that have finite buffer size (M/M/1/N), (N is the buffer size), models that include multiple servers (c servers), M/M/c/N, potentially a finite population of sensor nodes, where S sensor nodes are considered, namely M/M/c/N/S models, and finally general arrival and service models such as G/G/c/N. This can modify the limiting factors and can highlight different aspects that need to be improved to reduce the delay in the proposed architecture and can help further guide the processing allocation mode.

**Acknowledgements**
The authors would like to thank the Engineering and Physical Sciences Research Council (EPSRC), UK, for partly funding this work. This work was supported in part by the EPSRC INTERNET project under Grant EP/H040536/1, the EPSRC STAR Project under Grant EP/K016873/1 and the EPSRC TOWS Project under Grant EP/S016570/1. All data are provided in full in the results section of this paper.

**Amal A. Alahmadi** is currently a PhD student in the School of Electronic and Electrical Engineering, University of Leeds, UK. She is studying energy efficient vehicular, cloud, fog and optical networks. She has developed a number of new architectures for efficient processing in vehicular networks

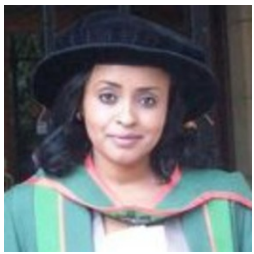

**Tasir E. H. EL-Gorashi** received the B.S. degree (first-class Hons.) in Electrical and Electronic Engineering from the University of Khartoum, Khartoum, Sudan, in 2004, the M.Sc. degree (with distinction) in Photonic and Communication Systems from the University of Wales, Swansea, UK, in 2005, and the PhD degree in Optical Networking from the University of Leeds, Leeds, UK, in 2010. She is currently a Lecturer in optical networks in the School of Electronic and Electrical Engineering, University of Leeds. Previously, she held a Postdoctoral Research post at the University of Leeds (2010– 2014), where she focused on the energy efficiency of optical networks investigating the use of renewable energy in core networks, green IP over WDM networks with datacenters, energy efficient physical topology design, energy efficiency of content distribution networks, distributed cloud computing, network virtualization and big data. In 2012, she was a BT Research Fellow, where she developed energy efficient hybrid wireless-optical broadband access networks and explored the dynamics of TV viewing behavior and program popularity. The energy efficiency techniques developed during her postdoctoral research contributed 3 out of the 8 carefully chosen core network energy efficiency improvement measures recommended by the GreenTouch consortium for every operator network worldwide. Her work led to several invited talks at GreenTouch, Bell Labs, Optical Network Design and Modelling conference, Optical Fiber Communications conference, International Conference on Computer Communications, EU Future Internet Assembly, IEEE Sustainable ICT Summit and IEEE 5G World Forum and collaboration with Nokia and Huawei.

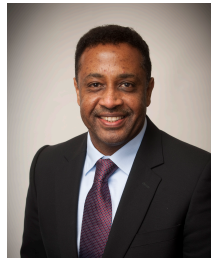

**Jaafar M. H. Elmirghani** is the Director of the Institute of Communication and Power Networks within the School of Electronic and Electrical Engineering, University of Leeds, UK. He joined Leeds in 2007 and prior to that (2000–2007) as chair in optical communications at the University of Wales Swansea he founded, developed and directed the Institute of Advanced Telecommunications and the Technium Digital (TD), a technology incubator/spin-off hub. He has provided outstanding leadership in a number of large research projects at the IAT and TD. He received the Ph.D. in the synchronization of optical systems and optical receiver design from the University of Huddersfield UK in 1994 and the DSc in Communication Systems and Networks from University of Leeds, UK, in 2012. He has co-authored Photonic switching Technology: Systems and Networks, (Wiley) and has published over 550 papers. He has research interests in optical systems and networks. Prof. Elmirghani is Fellow of the IET, Fellow of the Institute of Physics and Senior Member of IEEE. He was Chairman of IEEE Comsoc Transmission Access and Optical Systems technical committee and was Chairman of IEEE Comsoc Signal Processing and Communications Electronics technical committee, and an editor of IEEE Communications Magazine. He was founding Chair of the Advanced Signal Processing for Communication Symposium which started at IEEE GLOBECOM'99 and has continued since at every ICC and GLOBECOM. Prof. Elmirghani was also founding Chair of the first IEEE ICC/GLOBECOM optical symposium at GLOBECOM'00, the Future Photonic Network Technologies, Architectures and Protocols Symposium. He chaired this Symposium, which continues to date under different names. He was the founding chair of the first Green Track at ICC/GLOBECOM at GLOBECOM 2011, and is Chair of the IEEE Sustainable ICT Initiative, a pan IEEE Societies Initiative responsible for Green and Sustainable ICT activities across IEEE, 2012-present. He is and has been on the technical program committee of 41 IEEE ICC/GLOBECOM conferences between 1995 and 2021 including 19 times as Symposium Chair. He received the IEEE Communications Society Hal Sobol award, the IEEE Comsoc Chapter Achievement award for excellence in chapter activities (both in 2005), the University of Wales Swansea Outstanding Research Achievement Award, 2006, the IEEE Communications Society Signal Processing and Communication Electronics outstanding service award, 2009, a best paper award at IEEE ICC'2013, the IEEE Comsoc Transmission Access and Optical Systems outstanding Service award 2015 in recognition of "Leadership and Contributions to the Area of Green Communications", received the GreenTouch 1000x award in 2015 for "pioneering research contributions to the field of energy efficiency in telecommunications", the 2016 IET Optoelectronics Premium Award, shared with 6 GreenTouch innovators the 2016 Edison Award in the "Collective Disruption" Category for their work on the GreenMeter, an international competition, and received the IEEE Comsoc Transmission Access and Optical Systems outstanding Technical Achievement award 2020 in recognition of "Outstanding contributions to the energy efficiency of optical communication systems and networks", clear evidence of his seminal contributions to Green Communications which have a lasting impact on the environment (green) and society. He is currently an editor / associate editor of: IEEE Journal of Lightwave Technology, IEEE Communications Magazine, IET Optoelectronics, Journal of Optical Communications, and is Area Editor for IEEE Journal on Selected Areas in Communications (JSAC) Series on Machine Learning in Communication Networks (Area Editor). He was an editor of IEEE Communications Surveys and Tutorials and IEEE Journal on Selected Areas in Communications series on Green Communications and Networking. He was Co-Chair of the GreenTouch Wired, Core and Access Networks Working Group, an adviser to the Commonwealth Scholarship Commission, member of the Royal Society International Joint Projects Panel and member of the Engineering and Physical Sciences Research Council (EPSRC) College. He was Principal Investigator (PI) of the £6m EPSRC INTelligent Energy awaRe NETworks (INTERNET) Programme Grant, 2010-2016 and is currently PI of the £6.6m EPSRC Terabit Bidirectional Multi-user Optical Wireless System (TOWS) for 6G LiFi Programme Grant, 2019-2024. He has been awarded in excess of £30 million in grants to date from EPSRC, the EU and industry and has held prestigious fellowships funded by the Royal Society and by BT. He was an IEEE Comsoc Distinguished Lecturer 2013-2016.